# Slow-light Enhanced Correlated Photon-Pair Generation in Silicon

C. Xiong,[1,*] Christelle Monat,[1,2] Alex S. Clark,[1,3] Christian Grillet,[1] Graham D. Marshall,[4] M. J. Steel,[4] Juntao Li,[5] Liam O'Faolain,[5] Thomas F. Krauss,[5] John G. Rarity,[3] and Benjamin J. Eggleton[1]

[1]Centre for Ultrahigh-bandwidth Devices for Optical Systems (CUDOS), Institute of Photonics Optical Sciences (IPOS), School of Physics, University of Sydney, NSW 2006, Australia
[2]Institut des Nanotechnologies de Lyon, Ecole Centrale de Lyon, 36 Avenue Guy de Collongue, 69134 Ecully, France
[3]Centre for Quantum Photonics, University of Bristol, Queens Building, University Walk, Bristol, BS8 1TR, UK
[4]CUDOS, MQ Photonics Research Centre, Department of Physics & Astronomy, Macquarie University, NSW 2109, Australia
[5]School of Physics and Astronomy, University of St Andrews, Fife, KY16 9SS, UK
*Corresponding author: chunle@physics.usyd.edu.au

We report the generation of correlated photon pairs in the telecom C-band, at room temperature, from a dispersion-engineered silicon photonic crystal waveguide. The spontaneous four-wave mixing process producing the photon pairs is enhanced by slow-light propagation enabling an active device length of less than 100 µm. With a coincidence to accidental ratio of 12.8, at a pair generation rate of 0.006 per pulse, this ultra-compact photon pair source is immediately applicable towards scalable quantum information processing realized on-chip.

The integration of optical components for use in emerging quantum technologies such as quantum-secured communication [1] and quantum computation [2] is under intense investigation. A bright and low-noise single-photon source in the low-loss telecom bands is a key requirement for many applications. In this context, single quantum emitters have been exploited to realize solid-state single-photon sources [3, 4], but despite some recent advances, these approaches still present their own challenges, such as the need for low temperature, or the difficulty in integrating single emitters in practical devices. So far, the most common strategy to generate single photons relies on producing correlated photon pairs using a bulk optic nonlinear crystal. The photon pair generated by the underlying spontaneous nonlinear process allows for the detection of one photon heralding the arrival of the other. As nonlinear pair generation is stochastic, the photon states produced are non-deterministic and are often contaminated by states of higher photon number, reducing the fidelity of any quantum logic circuit using such states as inputs. Attenuating the power of the pump laser driving the nonlinear process reduces the fraction of higher number states, but at the expense of the source brightness. The development of a deterministic nonlinear single-photon source therefore requires many heralded sources that are pumped with sufficiently low power to preclude higher photon number states and that are multiplexed from a parallel architecture to guarantee the generation of single photons on command [5, 6]. To meet the demands of quantum information processing and its requirements of many single-photon input states and even cluster states [7, 8], a mature source may eventually contain hundreds or even thousands of individual pair generation units combined with intelligent routing to collect the photons into a single mode for delivery to any photonic application. To achieve these functionalities in a scalable scheme requires the miniaturization of each individual unit so that they can be integrated on a photonic chip.

Schemes investigated for such sources include spontaneous three-wave mixing using periodically poled lithium niobate (PPLN) waveguides [9], and spontaneous four-wave mixing (SFWM) using chalcogenide glass rib waveguides [10] or silicon nanowires [11–14]. In each case, the flux of generated single photons grows with the effective interaction strength $\gamma PL$, where $P$ is the pump power, $L$ the device length, and $\gamma$ measures the strength of the optical nonlinearity. The required path lengths of PPLN and chalcogenide waveguides are typically a few centimeters. The use of highly nonlinear silicon nanowires decreases the device length to ~ 1 cm, but further reducing the device size has remained a challenge because of the requirement for increased device nonlinearity.

Recently it has been reported that photonic crystal (PhC) structures can dramatically enhance the effective nonlinearity of silicon waveguides through slow-light propagation [15–18]. In particular, the advent of dispersion-engineered slow-light waveguides has enabled the demonstration of broadband stimulated four-wave mixing in devices of ~ 100 µm length [17, 18], two orders of magnitude shorter than silicon nanowires. In this letter, we report the generation of ~ 1550 nm correlated photon pairs in a low-loss silicon PhC waveguide which has been dispersion-engineered to exploit slow-light enhanced SFWM. This enhancement allows the length of the device to be reduced to 96 µm while retaining sufficient SFWM efficiency, making this the most compact emitter of correlated photon pairs yet reported.

The generation of correlated photon pairs through SFWM is illustrated in Fig. 1a. A coherent pulse of light enters the PhC region, where two photons from the pump are converted to signal and idler photons of higher and lower frequency respectively to form a quantum correlated state. Due to the slow-light effect of the PhC, the pump travels at a group velocity of ~ c / 30, and is strongly spatially compressed. Compared with silicon nanowires, this enhances the nonlinear interaction per unit device length per unit pump power approximately 100 fold.

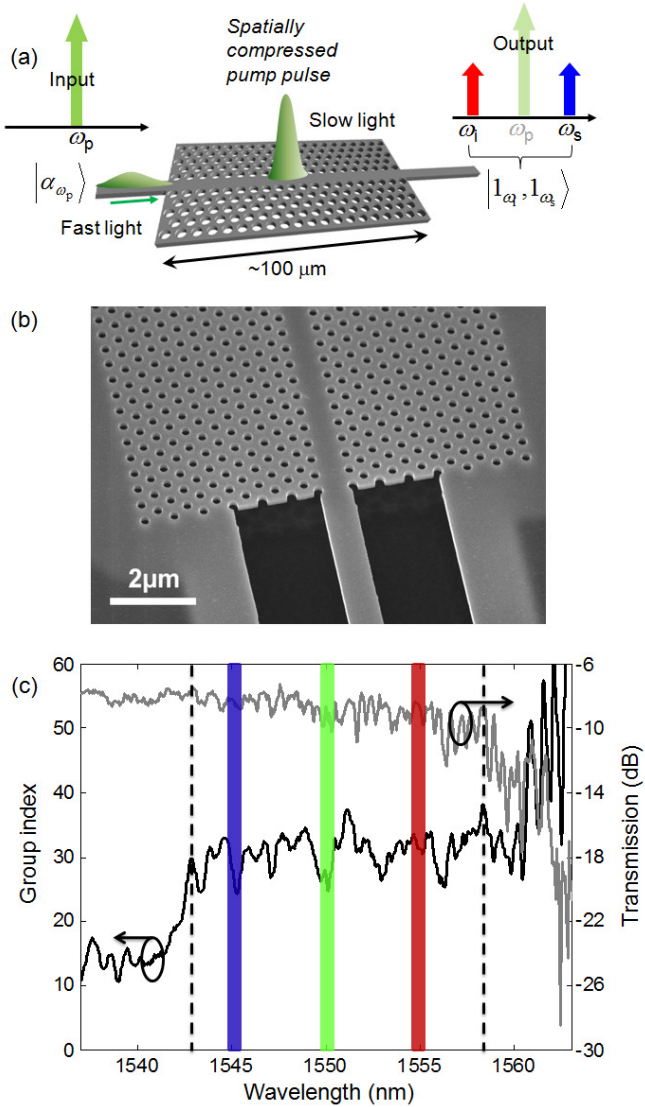

Fig. 1. (Color online) (a) Schematic of SFWM in a silicon slow-light PhC waveguide. (b) SEM image of the PhC waveguide. (c) The group index and total transmission of light in the silicon PhC waveguide. The window (between dotted lines) with a flat group index of ~ 30 and slightly increased loss defines the slow-light regime. The pump, signal and idler bands are represented by green, blue and red lines respectively.

The silicon slow-light device used in this study was fabricated on a silicon-on-insulator wafer comprising a 220 nm silicon layer on 2 μm of silica through the use of electron-beam lithography and reactive ion etching [17, 18]. The PhC waveguide (see Fig. 1b) was created from a triangular lattice of air holes etched in a suspended silicon membrane, with a row of holes missing along the ΓK direction. The rows adjacent to the waveguide centre were laterally shifted to engineer the waveguide dispersion such that it exhibits a group index of ~ 30 with low dispersion and moderate loss across a 15 nm window centered at 1553 nm (see Fig. 1c) [19]. Silicon access waveguides, including inverse tapers terminated by wide polymer waveguides were added to the input and output of the PhC region to improve coupling efficiency [18].

The PhC waveguide was probed as shown in Fig. 2. Alternate pulses from a 10 MHz repetition rate fiber laser provided a train of 1550.1 nm, 14 ps pulses at an effective rate of 5 MHz (the maximum trigger frequency of our detection system). The pulses passed through a filtering stage consisting of a 1 nm band-pass filter (BPF), a circulator and a fiber Bragg grating (FBG, 0.5 nm, reflection > 99%), and were TE polarized, before coupling to the chip using lensed fibers. Generated signal and idler photons at 4.8 nm detuning from the pump were postselected and separated using an arrayed waveguide grating (AWG), and filtered using 0.5 nm BPFs before being sent to InGaAs/InP single-photon detectors (SPDs, id-Quantique 201). The 4.8 nm detuning was chose so that the pair-photon wavelengths were sufficiently far from the pump for less leaked pump noise and were within the SFWM bandwidth (~ 15 nm) for efficient pair generation. The detection probability for both SPDs was set at 10% to reduce the dark counts. Taking into account the waveguide output coupling loss, the AWG and BPFs insertion loss, and the detection inefficiency, the total loss of each channel was estimated to be 21.8 dB. The simultaneous detection of a photon at both detectors was recorded as a coincidence ($C_{raw}$). We measured accidental coincidence counts ($A$) by changing the delay between detectors to trigger them on consecutive pulses where no photon pair correlation exists. The net coincidence count is $C = C_{raw} - A$. To improve measurement statistics, we acquired counts for 300 s.

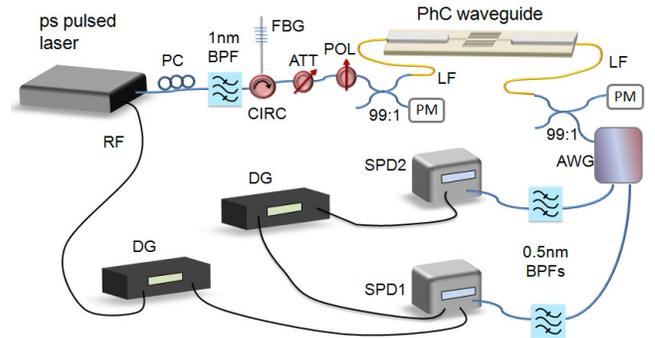

Fig. 2. (Color online) The experimental setup, RF: radio-frequency signals synchronized to laser pulses, PC: polarization controller, CIRC: circulator, ATT: attenuator, POL: polarizer, LF: lensed fiber, PM: power meter, DG: delay generator.

Figure 3 shows that the coincidences (blue diamonds) are consistently higher than the accidentals (blue circles), confirming the generation of correlated photons. It can be seen from Fig. 3 that the net coincidence (red squares) is quadratically dependent on the pump peak power ($P$) when $P < 0.42$ W, which is expected of the SFWM process. When $P \geq 0.42$ W, the net coincidence deviates from the quadratic fit (solid line) due to the onset of two-photon absorption (TPA) and the resulting free-carrier absorption (FCA) in silicon. Figure 3 also displays the measured coincidence to accidentals ratio (CAR = $C/A$, green triangles) obtained at various power levels. A maximum CAR of 12.8 is achieved for a coupled peak power of 0.23 W. We attribute the roll-off of the CAR beyond this power level to TPA and FCA. At the pump power level for the maximum CAR, the measured net coincidence is 383 per 300 s. After taking into account the total loss from the photon generation to detection of 21.8 dB in each channel, the intrinsic average pair generation rate is calculated to be 0.006 per pulse.

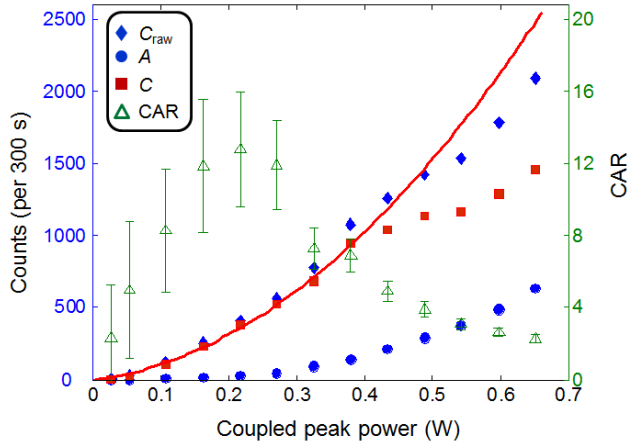

Fig. 3. (Color online) The measured coincidence ($C_{raw}$), accidental ($A$), net coincidence ($C$) counts in 300 s and CAR as a function of coupled pump peak power.

Ideally the CAR is proportional to the reciprocal of the average number of pairs per pulse if there is no noise such as pump leakage, Raman and detector dark counts [11, 13]. The upper limit of CAR for 0.006 pairs per pulse should be about 167. The silicon platform does not suffer from Raman noise [13], so the maximum CAR in our experiment was mainly limited by the pump leakage and detector dark counts. Improvements in future experiments will include inserting a FBG before the AWG to further suppress pump leakage, and increasing the pair photon to dark count ratio by optimizing waveguide output coupling efficiency and using low-loss filtering devices [13]. Figure 3 tells us that the photon to dark count ratio cannot be enhanced by increasing pump power because of the TPA and FCA effects in silicon.

In conclusion, we have demonstrated slow-light enhanced correlated photon-pair generation in an ultra-compact 96 μm long silicon PhC waveguide. In quantum information science applications such as quantum key distribution a source is considered to be useful when it has a CAR > 10. Our source exceeds this specification, achieving a CAR of 12.8 at a pair generation rate of 0.006 per pulse. Operating in this low pairs per pulse regime will allow the parallel generation of many photons without contamination from multiphoton effects, where efficient heralding and routing of generated photons facilitates the creation of a pseudo-deterministic single-photon source. This compact and integrated single photon source unit offers a scalable approach to quantum information processing on-chip.

This work was supported in part by the Centre of Excellence and Federation Fellowship programs of the Australian Research Council. The silicon waveguide chip was fabricated under the Engineering and Physical Sciences Research Council – UK silicon photonics consortium and supported by the European Union (EU) Seventh Framework Programme Marie Curie project "OSIRIS". ASC acknowledges the WUN Research Mobility Award and JGR acknowledges the Royal Society Wolfson Merit Award.